# The Reliability of X-ray Constraints of Intrinsic Cluster Shapes

David A. Buote[1]

Massachusetts Institute of Technology

John C. Tsai[2]

NASA/Ames Research Center

## ABSTRACT

Using the simulation of Katz & White (1993) we have tested the viability of X-ray analysis for constraining the intrinsic shapes of clusters of galaxies considering the effects of both substructure and steep temperature gradients. We restrict our analysis to the aggregate shapes of clusters on scales of $r \sim 1-2$ Mpc in order to reduce our sensitivity to subclustering in the core. For low redshifts ($z \lesssim 0.25$) the X-ray method accurately measures the true ellipticity of the three-dimensional cluster dark matter provided the inclination of the cluster is known to within $\sim 30°$; assuming the gas is isothermal adds only small errors to the derived shapes. At higher redshifts the X-ray method yields unreliable results since the gas does not trace the cluster gravitational potential. We proffer some necessary conditions for the reliability of X-ray methods characterized by both the amount of substructure in the X-ray surface brightness images and the shapes of the isophotes. We conclude that measurements of the aggregate shapes of clusters on scales $r \sim 1-2$ Mpc are insensitive to core substructure representing scales of a few hundred kpc. Therefore our results suggest that the X-ray measurements of aggregate cluster shapes by Fabricant, Rybicki, & Gorenstein (1984) and Buote & Canizares (1992) are valid provided that they do not suffer from serious projection effects. A substantial number of Abell clusters observed with the ROSAT PSPC will be amenable to X-ray shape analysis.

## 1. Introduction

The intrinsic shapes of galaxy clusters provide valuable information regarding the cosmological framework in which they were formed (e.g., Eisenstein & Loeb 1994). For example, "Pancake" theories (e.g., Sunyaev & Zeldovich 1972) for structure formation predict mostly oblate clusters while predominantly prolate structures result from theories invoking early tidal distortions (e.g., Binney & Silk 1979). The distribution of intrinsic cluster shapes also probes the nature of the dark matter itself. If, for example, the dark matter in clusters is generally rounder than the galaxy distribution then the dark matter must be dissipational to some extent (e.g., Strimple & Binney 1979; Aarseth & Binney 1978).

---

[1]dbuote@space.mit.edu

[2]jcht@cloud9.arc.nasa.gov

As first described in the pioneering papers by Binney & Strimple (1978; Strimple & Binney 1979) X-ray images of clusters provide a powerful probe of their intrinsic shapes. The shape of the three-dimensional cluster mass may be determined from a cluster X-ray image if the X-ray emitting gas is in hydrostatic equilibrium and an assumption is made regarding the cluster mass distribution along the line of sight. Note that these assumptions are not necessary for mapping the projected cluster mass with gravitational lens techniques that use the distorted images of background galaxies (Tyson, Valdes, & Wenk 1990; Kaiser 1992; Kaiser & Squires 1993; Fahlman et al. 1994; Smail et al. 1994). Because of the required source-lens-observer separations, however, these gravitational lens methods are generally restricted to high-redshift clusters ($0.15 \lesssim z \lesssim 0.6$). In contrast, X-ray methods in principle can be applied at any redshift.

The increasing evidence for substructure in clusters of galaxies suggests that many clusters are young and thus not dynamically relaxed (see Jones & Forman 1992; Bird 1993; West 1994). As a result, the reliability of methods that infer the intrinsic structure of clusters by assuming hydrostatic equilibrium of their X-ray emitting gas has been called into question (for a review see Fitchett 1989; also see Mohr, Fabricant, & Geller 1993). The particular case of A2256 dramatizes this uncertainty.

Fabricant, Rybicki, & Gorenstein (1984) and Fabricant, Kent, & Kurtz (1989) analyzed optical and *Einstein* X-ray data of A2256. Although they could not find any evidence for substructure in their data, they concluded that the observed elongation of the X-ray isophotes was consistent with the mass of Abell 2256 being either a single flattened spheroid ($\epsilon \sim 0.4$) or the superposition along the line of sight of two spherical masses separated by a few hundred kpc in the plane of the sky. The ROSAT image of A2256 (Briel et al. 1991) confirmed the latter assertion of substructure; Buote (1992) and Davis & Mushotzky (1993) re-analyzed the *Einstein* image and located the same substructure.

Does the subclustering on scales of a few hundred kpc invalidate the characterization of the shape of A2256 by a single flattened spheroid? If one is interested in measuring the aggregate shape of the cluster on scales of $1 - 2$ Mpc from the cluster center, and the shape is quantified by the quadrupole moment of the mass on those scales (or equivalently the principal moments of inertia), then subclustering on scales of a few hundred kpc represents higher order moments which should be unimportant with respect to the quadrupole term in the cluster potential. Hence, the aggregate cluster shape on scales of $1 - 2$ Mpc should be insensitive to small-scale subclustering.

Another limitation of using X-ray images to determine cluster shapes has been the lack of spatially resolved temperature profiles (e.g., Fitchett 1989). The predicted radial mass distribution depends strongly on the temperature gradient. However, Strimple & Binney (1979) and others (see §4.1.) have argued that conclusions about the shapes of clusters are not overly sensitive to temperature gradients.

Buote & Canizares (1992; hereafter BC92) analyzed the *Einstein* images of five low-redshift Abell clusters having no obvious subclustering (with the possible exception of Coma) in order to measure the aggregate shapes of the underlying mass on a scale $r \sim 1$ Mpc. They demonstrated that for all the clusters the shapes of the X-ray isophotes ($\epsilon_x \sim 0.15$) were rounder than the inferred dark matter distributions ($\epsilon_{DM} \sim 0.30$) which were rounder than the galaxy isopleths ($\epsilon_{gal} \sim 0.50$). In their analysis BC92 assumed hydrostatic equilibrium and isothermality for the X-ray emitting gas and that subclustering was dynamically unimportant for measuring the aggregate cluster shapes. It is our purpose to ascertain whether these assumptions are indeed justified.

Katz & White (1993; hereafter KW) modeled the formation and evolution of a Virgo-sized cluster ($M \sim 2 \times 10^{14} M_{\odot}$) in a standard flat, biased Cold Dark Matter universe ($\Omega = 1$, $H_0 = 50$ km s$^{-1}$ Mpc$^{-1}$, b=2.6, $M_{DM}/M_{bary} = 10$). They modeled the dissipational gas component with smoothed particle hydrodynamics (Hernquist & Katz 1989) allowing for cooling via radiative and compton processes; gravitational effects were modeled using the hierarchical tree method. Hence, KW constructed an X-ray cluster that formed and evolved in the context of large-scale structure and thus serves as a laboratory for testing X-ray methods for determining intrinsic cluster shapes. Using the KW cluster in its final time step Tsai, Katz, & Bertschinger (1994; hereafter TKB) tested the accuracy of spherically-symmetric X-ray analysis of the radial mass distribution. They concluded that the mass inferred from the X-rays matched the true mass of the simulation to within $\sim 25\%$ when the true temperature profile was used.

We will address how reliably the aggregate shape of the KW cluster on scales $r \sim 1 - 2$ Mpc can be determined from X-ray analysis as a function of redshift, and thereby assess the validity of the assumptions made by BC92. In §2. we discuss the surface brightness of the cluster; in §3. we discuss the intrinsic properties of the cluster obtained directly from the KW simulation; in §4. we discuss the X-ray modeling procedure and its results; in §5. we discuss the implications and in §6. we present our conclusions.

## 2. Analysis of the X-ray Surface Brightness

We created X-ray images by folding two-dimensional projections of the simulated cluster of KW through the instrument parameters of the *Einstein* Imaging Proportional Counter (IPC) (Giacconi et. al. 1979) to facilitate comparison to previous X-ray determinations of the shapes of the mass distributions in clusters of galaxies (Fabricant et al. 1984; Buote 1992; BC92). Specifically, the point spread function (PSF) and the energy response matrices of the IPC were applied to the X-rays from the KW simulation (see TKB). Although the IPC has only $\sim 1/3$ the spatial resolution of the ROSAT Positional Sensitive Proportional Counter (PSPC) and significantly less energy resolution, analysis of IPC-convolved cluster images is of equal importance to future studies of PSPC cluster images because (1) the cluster emission important for analysis extends over 50′ in diameter which renders the differences in the PSF's of the two instruments insignificant for analysis of the large-scale structure of the image, and (2) the spectrum of the image is not used for analysis of the shapes (although use of the exact temperature profile derived from the KW simulations by TKB will be employed for comparison to the temperature-blind analysis). Note that the IPC is also sensitive to energies (0.2 - 4 keV as opposed to 0.1 - 2.4 keV for the PSPC) that are more characteristic of temperatures of rich clusters.

We do not want the accuracy of our measurements to be limited by the skill of the observer or by noise. As a result, we allowed the image to be uniformly exposed with the on-axis effective area parameters of the IPC which corresponds to a perfect flat-field correction by the observer. Moreover, we allowed the image to be exposed for $2 \times 10^6$s which effectively eliminated statistical uncertainties; this exposure time translates to $5 - 10 \times 10^6$ counts for each image. Hence we constructed essentially perfect X-ray surface brightness maps.

Each X-ray image is a $120 \times 120$ field of 1′ square pixels. Following the convention of TKB we place the cluster at a fiducial distance of 100 Mpc; i.e. each pixel represents 29.1 kpc. We ignore the limited spectral information of the IPC-convolved images by adding all counts in PI bins 2-9. This translates to photon energies spanning the range 0.2-3.1 keV (note: BC92 used 0.2-3.5 keV).

Locating and removing any pockets of emission due to individual galaxies in the cluster is all that remains to prepare the images for analysis. This reduction procedure is essential since our hydrostatic analysis (see §4.) requires that the X-rays are due to thermal emission powered by the gas falling into the smooth underlying cluster gravitational potential. The artificial enhancement of the emission from regions of high gas density due to shortcomings in the KW simulation (also see TKB) exacerbates the contamination of the surface brightness map from local dips in the potential due to large galaxies. Fortunately these excess peaks are easily located by visual examination of the image enabling us to either avoid those regions or remove the excess. We describe in detail the reduction of the three projections of the $z = 0.13$ cluster and then summarize the results for higher redshifts.

## 2.1. Cluster at z=0.13

KW stopped their simulation at $z = 0.13$ when the evolution of the cluster had slowed substantially; in the core two globs (i.e. "galaxy-like objects") are merging while two other globs orbit about 25′ from the center. This redshift represents the cluster at its most relaxed state in the simulation and thus, of all the redshifts, best matches the criteria for X-ray analysis of cluster shapes outlined in §6 of BC92. This cluster, however, has a size and mass comparable to the Virgo cluster which is approximately an order of magnitude less massive than the rich Abell clusters studied by BC92. Moreover, the KW cluster underwent a major merger at $z \sim 0.6$. The clusters studied by BC92 may be, as a result, more evolved than this simulated cluster (this may not be true for Coma as Fitchett & Webster [1988] have argued for a recent merger).

KW project the three-dimensional cluster emission along three random orthogonal axes $x$, $y$, and $z$. In Figure 1 we show contour plots of X-ray images for each projection "observed" with the IPC as described above. Just outside the unrealistically over-dense core region, the emission from extended cluster gas dominates the surface brightness profile. Farther out, the emission from globs substantially distorts the surface brightness contours away from the contour shapes that would result from the smooth underlying cluster potential. Since in the outer regions the X-ray gas is significantly contaminated by individual globs and, moreover, is less dynamically relaxed, we restrict our analysis of the surface brightness to $r \lesssim 40' \sim 1.2$ Mpc of the cluster center. Within $40'$ two prominent sources in excess of the cluster continuum are seen in all three projections. In addition, the isophotes within $\sim 5'$ exhibit asymmetrical distortions due to the gravitational effects of the merging of two globs described by KW.

### 2.1.1. Source Removal

The satisfactory removal of sources embedded in a continuum is problematic when the goal is not to introduce spurious results into derived shape parameters of the continuum (small errors in the subtraction of such a source are unimportant when computing azimuthally-averaged quantities such as the radial profile). Effects resulting from incomplete removal of sources are especially serious when the continuum is nearly circular because any distortion will generate a preferred direction and a non-zero ellipticity. For example, complete subtraction of a source depends crucially on the accuracy of the modeling of the source and continuum. Smoothing the image biases the shapes of the continuum according to the functional form of the smoothing function.

We removed the embedded sources by instead exploiting the symmetry of the surface brightness distribution imposed by the dark matter models in §4.; i.e. the dark matter potentials generate X-ray isophotes that are concentric, of constant orientation, and possess two orthogonal axes of reflection symmetry. A given source, thus, may be replaced with the emission in the regions obtained by reflecting the source over the symmetry axes of the image. This type of "symmetric substitution" will not add any geometrical effects into the image that are not already being assumed (Strimple & Binney 1979) and is thus more reliable than direct subtraction of the source or smoothing the image. Unfortunately this method is limited to a continuum having a small number of embedded sources where there exists smooth continuum regions available for symmetric substitution for each source.

Consider first the central $40'$ of the $x$-projection. We determine the axes of symmetry by computing the ellipticity and position angle (see below) of the region uncontaminated by the individual sources. Specifically, this entails looking for (1) bumps in the radial profile (see below) and (2) sudden large changes in the ellipticity and position angle as a function of radius; we also exclude the inner $5'$ where the simulation becomes unphysical (see KW and TKB). The uncontaminated region lies between $5'$ and $19'$ and the position angle of the coordinate system belonging to the symmetry axes is rotated by $20°$ counter-clockwise from the image coordinate system; note that we do not attempt to replace the irregularly-shaped isophotes to the right of the center since they are not well localized. In terms of this new coordinate frame, the two obvious sources within $40'$ lie in the first and second quadrants. We replace the source in the first quadrant (a circular region of radius $7.5'$) with the average emission of the reflection-related regions in the third and fourth quadrants; the corresponding region in the second quadrant is contaminated by the other source. Similarly, the source in the second quadrant (a circular region of radius $12.5'$) is replaced by the average emission of the corresponding reflection-related regions in the third and fourth quadrants. We followed precisely the same procedure to remove the sources in the $y$ projection with the exceptions that a smaller region from $5' - 13'$ was identified as the uncontaminated region, the corresponding coordinate system of the symmetry axes is rotated by $10°$ counter-clockwise with respect to the image coordinates, and we also removed the source in the lower right of the image ( see Figure 1). We show the images of the $x$ and $y$ projections with sources removed by symmetric substitution in Figure 2.

The $z$ projection requires more care than the $x$ and $y$ projections because the isophotes are nearly circular away from the two sources in the first quadrant. We make the conservative assumption that the effects of the two sources are contained within half of the image and thus may be removed by replacing the contaminated half with the other half of the image. Unfortunately, the emission from the two sources bias the choice of reflection axis practically to the region where the unphysical central emission dominates. Thus we examine the effects of several different reflection axes oriented by angles spanning $90°$ through $180°$. We find that the new position angles of the isophotes are always aligned with the reflection axis and the ellipticities vary between $0.00 - 0.12$ (for semi-major axes between $5'$ and $20'$). As there is no priori reason to discriminate between any of these symmetry axes we have to accept that only an upper limit on the ellipticity of $0.12$ can be reliably set for the $z$ projection. However, for ensuing analysis we use $110°$ as the rotation angle for the symmetry axis because the isophotes appear the most regular in that case. We display this corrected image for the $z$ projection in Figure 2.

### 2.1.2. Radial Profile

Following Buote & Canizares (1994, hereafter BC94) we constructed azimuthally-averaged radial profiles for each image in radial bins of $1'$ width. The centroid of each profile was determined by the origin of the symmetry axes described above and is consistent to better than $1\%$ with the centroid computed from the symmetrically-substituted image for different values of the outer radii. We plot the radial profiles for the three projections in Figure 3. In order to reduce contamination from gas that is not completely relaxed, we restrict analysis to $r \leq 40'$ for each image. We exclude the central $5'$ from analysis because of the unphysical behavior of the simulation; even if the KW simulation produced a realistic cooling flow, we would ignore that region since our modeling procedure does not incorporate multi-phase structure in the gas (see §4.).

Historically the X-ray surface brightness ($\Sigma_X(r)$) of clusters has been parametrized by the $\beta$-model (Cavaliere & Fusco-Femiano 1976; Jones & Forman 1984; Sarazin 1986),

$$\Sigma_X(r) \propto \left[1 + \left(\frac{r}{a_X}\right)^2\right]^{-3\beta+1/2},$$  (1)

where $a_X$ and $\beta$ are free parameters. We determine the $\beta$-model parameters for each image in order to provide a convenient benchmark to compare to real clusters. In order to obtain physical constraints on these parameters the $\beta$-model must be convolved with the PSF of the IPC. The IPC is well approximated by a Gaussian of $1.5'$ FWHM over our energy range (Fabricant, Lecar, & Gorenstein 1980; Trinchieri, Fabbiano, & Canizares 1986). Since the cluster image is much larger than the scale of the PSF, folding the PSF into the $\beta$-model actually has a very small effect on the fits; nevertheless we do include it in the fits for completeness. We plot the best-fit $\beta$-model and list the best-fit parameters of the images in Figure 3. The quality of the fits is formally atrocious due to the minuscule statistical uncertainties of the "observation". However, the fits are visually outstanding. Our results differ slightly from TKB due to the different bins included in the fits; i.e. TKB include two bins at smaller $r$ and and we include ten bins at larger $r$. Note that the quality of the fit of the $\beta$-model to this simulated cluster far surpasses the fit to the IPC image of Coma but is of similar quality to that of A2256 (see Davis & Mushotzky 1993). Both Coma and A2256 have significant core substructure evident from their IPC and PSPC images (Briel et al. 1991; Buote 1992; Davis & Mushotzky 1993; Mohr, Fabricant, & Geller 1993). Hence this simulated cluster (at $z = 0.13$) may be related to the class of clusters having core substructure and good fits to the $\beta$-model like A2256. Clusters like Coma that are not well fit by the $\beta$-model may have significantly different structure from this cluster (at $z = 0.13$).

### 2.1.3. Ellipticity

As described in BC94, we quantify the shape of the X-ray surface brightness using the iterative moment technique introduced by Carter & Metcalfe (1980) to measure the shapes of the galaxy isopleths in rich clusters. In essence this technique entails computing the two-dimensional principal moments of inertia in an elliptical region arrived at by iterating an initially circular region; the square root of the ratio of principal moments yields the axial ratio and the orientation of the principal moments yields the position angle. The parameters obtained from this method, $\epsilon_M$ and $\theta_M$, are good estimates of the ellipticity ($\epsilon$) and the position angle ($\theta$) of an intrinsic elliptical distribution of constant shape and orientation. For a more complex distribution, $\epsilon_M$ and $\theta_M$ are average values weighted heavily by the outer parts of the region. This property is especially desirable because the primary objective of this paper is to measure the underlying smooth shape of the mass distribution, and visual examination of the images (Figure 2) clearly shows that the isophotes are far from perfect ellipses. Standard elliptical isophote fitting, in contrast, would emphasize the local irregularities at each radius instead of a gross value for a large annular region.

We compute $\epsilon_M$ and $\theta_M$ in an elliptical aperture for several values of the semi-major axis $a$ omitting the inner $5'$; the derived shape parameters are actually not sensitive to the peculiarities of the inner regions since the moments are heavily weighted by the outermost parts of the aperture. For each image we list $\epsilon_M$ and $\theta_M$ as a function of aperture semi-major axis in Table 1. Due to the extremely large number of counts the statistical and systematic uncertainties (see BC94; Carter & Metcalfe 1980) are negligible for these parameters. Performing Monte Carlo simulations for simulated clusters having the characteristics of each image (i.e. $\epsilon_M$, $\beta$, and $a_X$) we estimate the 99% confidence limits to be $\Delta\epsilon_M < 0.005$ and $\Delta\theta_M < 1°$ for the regions listed in Table 1.

The ellipticities are not a strong function of aperture size: $\epsilon_M \sim 0.27$ in the $x$-projection, $\epsilon_M \sim 0.15$ in the $y$-projection, and $\epsilon_M \sim 0.07$ in the $z$-projection. For a given projection, the differences in $\epsilon_M$ for different semi-major axes, although small ($\lesssim 0.05$), are significant. Similarly the position angle variations are small ($\lesssim 10°$), but significant, in these regions. These deviations reflect either the perturbations of the surface brightness resulting from the gravity of individual globs (i.e. substructure), oscillations of the gas due to incomplete relaxation, or any other intrinsic variations of the smooth underlying potential not accounted for in the above Monte Carlo simulations. We mention that the centroids of these regions are consistent within their uncertainties determined by the above Monte Carlo simulations.

The ellipticities of these images are consistent with those obtained by BC92 with the IPC for A401, Perseus, Coma, A2029, and A2199 (see also McMillan, Kowalski, & Ulmer 1989; Mohr, Fabricant,& Geller 1993; Davis & Mushotzky 1993). The agreement is better for the ellipticities of the $y$ and $z$ projections since the $x$-projection ellipticity is larger by $\sim 0.10$. However, BC92 argue that the clusters they analyze are probably viewed in nearly their most flattened projection (i.e. edge-on) suggesting that the $x$ and perhaps $y$ projections should be compared to the BC92 clusters (we discuss the intrinsic orientation of the dark matter in §3.). The $x$-projection has an ellipticity more similar to that of A2256 where $\epsilon \sim 0.25$ within 1 Mpc (Fabricant et al. 1984; McMillan et al. 1989; Buote 1992; Mohr et al. 1993; Davis & Mushotzky 1993). This, when coupled with the good fit of the $\beta$-model to A2256 (see above), implies that the shape and radial structure of this cluster is similar to A2256, a cluster likely less relaxed than the clusters studied by BC92 (with the possible exception of Coma).

## 2.2. Cluster at higher redshifts

We analyzed the surface brightness of the cluster at higher redshifts using the same techniques as for the $z = 0.13$ case. For the sake of brevity we restrict our attention to the $x$-projection since that is where the cluster is nearly in its most flattened state; i.e. we are only interested in assessing the accuracy of the dark matter shape determinations as a function of redshift, not as a result of projection effects. We show in Figure 4 the surface brightness maps at $z = 0.83, 0.67, 0.38$, and $0.25$.

The emission of the $z = 0.83, 0.67$, and $0.38$ clusters is punctuated by several peaks (due to individual glob emission). As a result it is difficult to obtain a meaningful value for the ellipticity anywhere for $z = 0.83$ and $z = 0.67$, although for $z = 0.38$ the ellipticities should be reliable for $r > 25'$. Excluding the region interior to $25'$ and exterior to $40'$ we obtain $\epsilon \sim 0.5$ for $z = 0.83$ and $\epsilon \sim 0.11$ for $z = 0.38$. For $z = 0.67$ we exclude the region interior to $25'$ (and hold the center fixed) and obtain $\epsilon \sim 0.4$ for outer radius $r \sim 40'$. We can remove the glob emission from the radial profiles of the surface brightness by simply flagging the affected pixels and excluding them from analysis. The $z = 0.67$ and $z = 0.38$ clusters are fit well by the $\beta$-model for $r = 0.2 - 1$ Mpc, but the $\beta$-model is not a good description of the surface brightness of the $z = 0.83$ cluster.

For the $z = 0.25$ cluster we remove two sources by symmetric substitution as done for the $z = 0.13$ cluster; we show the result in Figure 5. The isophotes of the $z = 0.25$ cluster are quite similar in shape to those of the $z = 0.13$ cluster, but are slightly rounder; we list the ellipticities in Table 1. Like the $z = 0.13$ cluster the $\beta$-model fits the surface brightness well, although the profile is somewhat flatter than for $z = 0.13$; we show the radial profile and the associated best-fit $\beta$ model in Figure 5.

The ellipticities of the $z = 0.83$ and $z = 0.67$ clusters are much larger than those of the clusters studied by BC92. These ellipticities are more typical of clusters having substantial substructure, to a degree not observed in the clusters of BC92; e.g., the double cluster A754 (Fabricant et al. 1986) and other bimodal clusters (see Jones & Forman 1991). Although within $25' < r < 40'$ the $z = 0.38$ cluster has a similar shape to the BC92 clusters, outside the cluster there is an elongated tail in the upper right (see Figure 4) and several obvious clumps ($r \lesssim 20'$) indicating substantially more structure. In contrast, the $z = 0.25$ cluster for $r \lesssim 40'$ is quite similar to the clusters of BC92 and should, along with the $z = 0.13$ cluster, provide a fair comparison of the reliability of shape determinations (§4.) to the BC92 clusters.

## 3. True Structure of the Dark Matter and Gas

In this section we summarize the three-dimensional structural properties of the cluster taken directly from the KW simulation; we refer the reader to KW for views of the cluster along the three orthogonal directions at different redshifts. Following the format of the previous section, we will discuss the $z = 0.13$ cluster at some length and briefly summarize the results for the other redshifts.

We compute the shape and orientation of the dark matter and gas particles following Katz (1991) who generalizes to three dimensions the two-dimensional iterative moment technique of

§2.1.. This method gives the shape and orientation for the particles within the ellipsoidal radius defined by

$$a^2 = x_1^2 + \frac{x_2^2}{q_{21}^2} + \frac{x_3^2}{q_{31}^2}, \tag{2}$$

where the $x_i$ refer to the coordinate system where the three-dimensional moment of inertia tensor is diagonal, $x_1$ is along the direction of the longest axis (i.e. largest principal moment), and the axial ratios $q_{i1}$ are the square roots of the ratios of principal moments of inertia in the $i$ direction to the 1 direction; the principal directions with respect to the fiducial $x - y - z$ coordinate system of KW give the orientation of the ellipsoid. Because of the finite number of particles for both the dark matter and gas, the ellipticities, $\epsilon_{i1} = 1 - q_{i1}$, and orientations (given by the Euler angles $\theta_i$ and $\phi_i$) are uncertain due to realization dependency (Strimple & Binney 1979). Following Strimple & Binney we construct pseudo-clusters having the same number of particles as the KW cluster and estimate the uncertainties of the derived shape parameters.

We wish to compute the aggregate shape of the underlying dark matter and gas, not the shape represented by the globs. The globs do not seriously bias the dark matter shapes, but they do substantially affect the gas shapes. Hence we need to remove the clumps via the symmetric-substitution procedure discussed in the previous section. This is important for the gas at distances $\sim 30'$ from the center.

In Table 2 we list for the dark matter and gas $\epsilon_{i1}$ and the relative orientations of the smallest principal moment $I_3$: $\theta_{DM}$ is the relative position of $I_3$ for the dark matter with respect to its value at $10'$; $\theta_{gas}$ is the relative position of $I_3$ for the gas and dark matter at a given radius; we omit the inner $5'$ for determining the gas shapes because of the unphysical behavior there. The uncertainties due to realization dependency are typically $\Delta\epsilon \sim 0.03$ and $\theta \sim 7°$. In Figure 6 we plot isodensity contours of the dark matter projected along the second-longest principal axis which thus corresponds to the most elongated projection (this principal axis is $\sim 10°$ from the $x$-projection – see below); note the projected dark matter in Figure 6 is smoothed with a Gaussian filter ($3'$ FWHM) for visual clarity.

The dark matter is nearly oblate for $r \lesssim 750$ kpc ($\sim 25'$) and becomes increasingly triaxial at larger radii. In a similar manner, the dark matter is rounder near the center ($\epsilon \sim 0.35$) and becomes increasingly elongated with distance. At $\sim 1.5$ Mpc ($\sim 50'$) from the cluster center, the dark matter ellipticity is 0.54 which is the fiducial value we will use to parametrize the aggregate shape of the cluster for comparison to the models in §4.. The orientation of the short axis, which we will call the "symmetry axis" since the cluster is nearly oblate, is nearly constant to within $\sim 10°$ at all radii. Taking the inclination angle $i$ to be defined with respect to the line of sight and the symmetry axis, we have $i = 80°$ for the $x$-axis, $i = 60°$ for $y$, and $i = 40°$ for $z$. Hence the $x$-projection is nearly edge-on, the $z$-projection largely face-on, and the $y$-direction intermediate.

The gas, like the dark matter, is nearly oblate within $\sim 750$ kpc and becomes increasingly triaxial farther out; the small $\epsilon_{21}$ values are consistent with those of the dark matter to within the stated uncertainties. In contrast, the values of $\epsilon_{31}$ are systematically less than those of the dark matter by $\sim 0.10$. In addition, the orientation of the symmetry axis of the gas is consistent with that of the dark matter to within $\sim 10°$. Both of these characteristics are consistent with the gas being in quasi-hydrostatic equilibrium with the smooth underlying potential of the dark matter and gas. Moreover, the isopotential shapes are consistent with those of the gas to within $\epsilon \sim 0.05$ and $\theta \sim 10°$; a complete exploration of the dynamical state of the gas will be done by others (N. Katz and A. Babul 1994, in preparation).

To quantify the radial structure of the dark matter and gas we binned each component into ellipsoidal bins using the above average shape properties. We parametrized the dark matter and gas densities by fitting power-law (see equation [15] of BC94) and Dehnen (1993) functions $(a^{-n}(a + a_c)^{n-4}, n = 1-4)$ to these density profiles. The power-law density yields good fits for $r > 50$ kpc but is too shallow in the core: $\rho_{DM} \sim r^{-2.2}$ with $r_c \sim 50$ kpc, $\rho_{gas} \sim r^{-2.2}$ with $r_c \sim 250$ kpc; i.e. the dark matter and gas have similar radial dependences, but the dark matter is more centrally condensed. The Dehnen function actually yields an excellent fit to the dark matter all the way into the core and follows very nearly the Hernquist (1990) form (i.e. $a^{-1}(a + a_c)^{-3}$).

The ellipticity profiles of the dark matter for the higher redshifts $z = 0.83, 0.67, 0.38$, and $0.25$ do not differ much from that of $z = 0.13$. The higher redshifts are slightly more elongated reaching a maximum $\epsilon_{DM} = 0.60$ for $z = 0.83$ computed within $\sim 1.5$ Mpc $(50')$ of the cluster center. Unfortunately, the number of globs at these higher redshifts (except for $z = 0.25$) substantially exceeds that at $z = 0.13$ and thus the gas ellipticities are severely contaminated by the glob emission. Nevertheless, we estimate for $z = 0.83$ that the gas, like the X-ray isophotes, is highly elongated ($\epsilon_{gas} \sim 0.5$) with nearly the same shape as the dark matter itself. The gas isodensity surfaces of the $z = 0.67$ cluster ($\epsilon_{gas} = 0.4$) are somewhat rounder than the dark matter but still flatter than the potential ($\epsilon_\Phi \sim 0.30$). For the $z = 0.38$ cluster the shape of the gas, like the X-ray isophotes, varies drastically with radius ranging from $\epsilon = 0.1 - 0.4$. For $z = 0.25$, in contrast, the gas is everywhere rounder than the dark matter and in fact traces the potential quite accurately, just like the $z = 0.13$ case.

The temperature gradients are similar for all of the redshifts considered (see Figure 11 of KW). The temperature outside of 50 kpc follows a steep negative gradient having $\frac{d\ln T_{gas}}{d\ln r} \sim -2.5$ while inside it falls rapidly due to the cooling flow (KW; TKB). In §5. we discuss the importance of this steep temperature gradient.

## 4. The Shape of the Three-Dimensional Dark Matter Distribution Deduced from the X-ray Images

### 4.1. Method

The technique we employ to constrain the shape of the cluster potential, and hence its mass, from the X-ray images derives from the pioneering work of Binney & Strimple (1978 Strimple &

Binney 1979) and is discussed in detail by BC94 (also BC92). The fundamental assumptions of this method are that the gas is a single-phase ideal gas in a state of quasi-hydrostatic equilibrium with the gravitational potential of the cluster. It would be relatively simple to incorporate effects due to rotation or a multi-phase structure of the gas, but since previous X-ray detectors have not been sensitive enough to place detailed constraints on such properties, the simplist conditions have been assumed. In particular, the inability of previous X-ray satellites to accurately measure the two-dimensional temperature profile of the gas restricts our ability to measure the gravitational potential. If the two-dimensional temperature of the gas is known precisely, then the best model-independent procedure to constrain the shape of the potential is to solve the hydrostatic equation for the potential in terms of the gas density and temperature, both of which may be determined from direct deprojection of the surface brightness and the spatially-resolved spectra (e.g., using the Lucy deprojection algorithm applied to spheroidal systems by Binney, Davies, & Illingworth 1990). Since, however, the temperature profile is generally poorly known, a more practical approach to constrain the unknown potential is to exploit the best-determined quantity, the surface brightness, while making "reasonable" assumptions about the temperature profile. As in previous studies (BC92; BC94) we adopt this approach by solving the equation of hydrostatic equilibrium for the gas density while assuming functional forms for the gas temperature and the gravitational potential. From the gas density we construct the X-ray emissivity and then, by projection onto the sky, the X-ray surface brightness. Finally, we convolve the surface brightness with the IPC PSF to compare to the "observed" images.

The first step in our analysis is to model the cluster gravitational potential. For simplicity we restrict the models to oblate and prolate spheroids in order to bracket the behavior of the general triaxial models (Strimple & Binney 1979). We consider the following two families of potentials: (1) potentials generated by mass distributions stratified on concentric, similar spheroids and (2) potentials which are themselves stratified on concentric, similar spheroids. Following the convention of Kassiola & Kovner (1993), who study the properties of two-dimensional elliptical potentials, we refer to three-dimensional potentials of model (1) as SMD (Spheroidal Mass Distributions) and (2) as SP (Spheroidal Potentials). Although the SP models have some properties that are undesirable for a physical mass model (see below), the constant shape of the potential and the ellipticity gradient of the mass distribution contrast nicely with the SMD's. (In addition, the simple analytic forms for the potential significantly increase computational speed.) Hence studying both SMD's and SP's allows for testing a wide range of cluster mass distributions which hopefully bracket the physical behavior of the real cluster.

The SMD potentials are generated by mass densities $\rho(m)$, where $m^2 = R^2/a^2 + z^2/b^2$, $R$ and $z$ are the conventional cylindrical coordinates, $a$ is the semi-major axis and $b$ the semi-minor axis of the spheroid that bounds the mass; full accounts of SMD potentials are given by Chandrasekhar (1969) and Binney & Tremaine (1987). As described in BC94, we consider mass densities having either a Ferrers (i.e. power-law) or Hernquist (1990) form. The free parameters of the SMD models are the core parameter, $R_c$, semi-major axis length, $a$, the ellipticity $\epsilon = 1 - b/a$ of the mass, and the power-law index $n$. Generally we fix $a$ and $n$ and normalize the potential to its central value. In this manner we construct potentials of varying scale ($R_c$) and shape ($\epsilon$).

The SP models are given by $\Phi = \Phi(\xi)$, where $\xi^2 = R^2 + z^2/q_\Phi^2$; $q_\Phi$ is the constant axial ratio of the SP such that $q_\Phi < 1$ for oblate and $q_\Phi > 1$ for prolate SP's. In particular, we consider the spheroidal logarithmic potential of Binney (1981; Binney & Tremaine 1987; also Kuijken & Dubinski 1994),

$$\Phi\left(R, z\right) = \frac{v_c^2}{2} \log\left(\frac{R_c^2 + \xi^2}{R_{ref}^2}\right), \tag{3}$$

where $v_c$ is the circular velocity $\sqrt{R d\Phi/dR}$ evaluated at infinity, $R_c$ is a core parameter of the potential, and $R_{ref}$ defines the unit of distance; in order that $\Phi$ not be positive we define $R_{ref}$ so that $R_c^2 + \xi^2 \leq R_{ref}^2$ for all $(R, z)$ considered. The mass density that generates this potential is,

$$\rho\left(R, z\right) = \left(\frac{v_c^2}{4\pi G q_\Phi^2}\right) \frac{\left(2q_\Phi^2 + 1\right) R_c^2 + R^2 + 2\left(1 - 1/2q_\Phi^2\right) z^2}{\left(R_c^2 + R^2 + z^2/q_\Phi^2\right)^2}. \tag{4}$$

Extending the Binney SP to a general power law we also consider the power-law potentials (Evans 1994; in two dimensions called "Tilted Plummer" models by Kassiola & Kovner 1993),

$$\Phi\left(R, z\right) = -\left(\frac{v_c^2(R_c, 0) 2^n R_c^{2n}}{n}\right) \left(R_c^2 + \xi^2\right)^{-n}, \tag{5}$$

where $n > 0$ and $v_c(R_c, 0)$ is the circular velocity evaluated at $(R_c, 0)$. The corresponding density is,

$$\rho\left(R, z\right) = \left(\frac{v_c^2(R_c, 0) 2^{n+1} R_c^{2n}}{4\pi G q_\Phi^2}\right) \frac{\left(2q_\Phi^2 + 1\right) R_c^2 + \left(1 - 2nq_\Phi^2\right) R^2 + 2\left(1 - \left(1 + 2n\right)/2q_\Phi^2\right) z^2}{\left(R_c^2 + R^2 + z^2/q_\Phi^2\right)^{n+2}}, \tag{6}$$

For particular values of $q_\Phi$ and $n$ the mass densities have peculiar properties; namely, the density can become "peanut-shaped" and possibly somewhere take negative values. These undesirable properties result because of the constant shape of the SP's. That is, the shape of the mass must counteract the tendency for the potential to become rounder with distance due to the rapid decay of higher order multipole moments. The density of the Binney potential, for example, has negative values on the $z$-axis for $q_\Phi < 0.707$ (Binney & Tremaine 1987). The power-law SP's have negative values somewhere on the $R$-axis when $q > 1/\sqrt{2n}$ ($n > 0$) and on the $z$-axis when $q < \sqrt{n + 1/2}$ ($n > -1/2$). The free parameters for these models are $R_c$, $\epsilon_\Phi$ (which is $1 - q_\Phi$ for the oblate models and $1 - 1/q_\Phi$ for prolate models), and $n$ for the power-law models; $R_{ref}$ in the logarithmic potential is arbitrarily fixed to the outer boundary of the X-ray gas. As with the SMD's we fix $n$ and normalize the potential to its central value $\Phi_0$; note we relate $\rho$ to $\Phi$ by $v_c^2(R_c, 0) = -n\Phi_0$. We then compute potentials of varying scale ($R_c$) and shape ($\epsilon_\Phi$).

The gas density is computed by assuming the gas is ideal and in quasi-hydrostatic equilibrium with the cluster potential; by "quasi" we mean that additional gas motions are dynamically unimportant with respect to the cluster gravity. TKB showed that the spherically-averaged $z = 0.13$ cluster is indeed in quasi-hydrostatic equilibrium; for purposes of shape analysis, however, this has not been demonstrated, although the agreement of the three dimensional shapes of the gas and potential in §3. is suggestive of hydrostatic equilibrium. In this paper we will assume that quasi-hydrostatic equilibrium holds for all $z$ and remark when the assumption appears to yield erroneous mass shape determinations.

If the gas is isothermal, the equation of hydrostatic equilibrium gives for the gas density,

$$\tilde{\rho}_{gas}(\vec{x}) = \exp\left[1 - \tilde{\Phi}(\vec{x})\right]^{\Gamma}, \tag{7}$$

where $\tilde{\rho}_{gas}$ and $\tilde{\Phi}$ are the gas density and potential normalized to their central values, $\Gamma = \frac{\mu m_p \Phi_0}{k_B T_{gas}}$, $\mu$ is the mean atomic weight, $m_p$ is the proton mass, $T_{gas}$ is the constant gas temperature, and $k_B$ is Boltzmann's constant. For a given potential $\Phi$, $\Gamma$ is well constrained by the radial profile of the X-ray surface brightness (see BC92 and BC94) and hence does not require knowledge of either the gas temperature ($T_{gas}$) or the potential depth ($\Phi_0$) (i.e. the total mass of the cluster). This simple solution is of particular interest for study of the cluster shape since detailed two-dimensional temperature maps were beyond the capabilities of past X-ray satellites. As first shown by Strimple & Binney (1979) and then by Fabricant et al. (1984), the shapes of the X-ray isophotes for a given $\Phi$ do not radically differ if the gas is assumed to be isothermal or adiabatic. BC94 generalized these findings by demonstrating that for a wide class of potentials and emissivities the shapes of the X-ray isophotes are very similar ($\Delta\epsilon \lesssim 0.04$), independent of the temperature gradient. Thus, the isothermal solution should yield an accurate estimate of the shape of the cluster even if the gas is not isothermal. The constraints on the radial distribution, however, will be in error for large temperature gradients. Since the $z = 0.13$ cluster has a steep temperature gradient, we have a formidable test of this assertion.

Since we know the exact temperature distribution for the cluster (§3.) we also consider the solution of the hydrostatic equation for an arbitrary temperature profile,

$$\tilde{\rho}_{gas}(\vec{x}) = \frac{1}{\tilde{T}_{gas}(\vec{x})} \exp\left[-\Gamma_0 \int_0^{\vec{x}} \frac{\nabla\tilde{\Phi}(\vec{x}') \cdot d\vec{x}'}{\tilde{T}_{gas}(\vec{x}')}\right], \tag{8}$$

where $\tilde{T}$ is the gas temperature expressed in terms of its value at $\vec{x} = 0$; the integral is independent of path. $\Gamma_0$ is equal to $\Gamma$ as given above for the isothermal solution except with $T_{gas}$ replaced by $T_{gas}(0)$.

We construct the X-ray emissivity $j_{gas}$ from $\rho_{gas}$ via the relation, $j_{gas} \propto \rho_{gas}^2 \Lambda_{IPC}(T_{gas})$, where $\Lambda_{IPC}$ is the plasma emissivity convolved with the IPC spectral response in the appropriate energy band (0.2 - 3.1 keV). Since $\Lambda_{IPC}$ is a relatively weak function of temperature (e.g., Fabricant et al. 1980; Trinchieri et al. 1986), we use the approximation $j_{gas} \propto \rho_{gas}^2$ for the isothermal models; for completeness we use the values of $\Lambda_{IPC}$ from Raymond & Smith (1977, updated to the current version) for the exact temperature models although the results hardly differ if we assume $\Lambda_{IPC}$ is constant. By integrating $j_{gas}$ along the line of sight we obtain the model X-ray surface brightness $\Sigma_X$. The final step to prepare the model image for comparison to observations is to convolve $\Sigma_X$ with the IPC PSF.

For each model we consider the cases $i = 90°$ (i.e. edge-on) and $i$ set to the true inclination angle of the symmetry axis of the cluster spheroid. We expect the errors in assuming the cluster to be edge-on to be substantial for large tip angles (i.e. $i \ll 90°$) because we can only measure the elongation of the cluster projected onto the plane of the sky; i.e. overestimating the inclination angle is equivalent to underestimating the intrinsic elongation of the cluster. Binney & Strimple (1978) and Fabricant et al. (1984) have shown that for moderate tip angles ($70 \leq i \leq 90$) the inferred shape of the underlying mass distribution is little affected. The $y$ and $z$ projections of the KW cluster at $z = 0.13$ have substantial tip angels and thus provide a test of these assertions.

## 4.2. Results for $z = 0.13$ Cluster

We determine the intrinsic shape of the underlying cluster mass by comparing $\epsilon_M$ and the azimuthally averaged radial profiles of the simulated X-ray images (§2.) to those generated by the models (§4.1.). Because we are concerned only with the aggregate shape of the cluster on scales $\sim 1.5$ Mpc from the center, not with small-scale perturbations due to individual globs, we quantify the elongation of the surface brightness by using the values of $\epsilon_M$ computed within $5' - 20'$ and $5' - 40'$; we consider the two values to accommodate a possible change in elongation with radius. Employing more values of $\epsilon_M$ at different radii for comparison would sample the cluster on scales smaller than those we are attempting to quantify with our aggregate analysis. For the same reason, we use the azimuthally averaged radial profile instead of fitting to individual pixels of the surface brightness. We could use elliptical annuli that better correspond to the shapes of the X-ray isophotes, but the fitted parameters and the quality of the fits is not affected.

For the SMD models we begin by specifying the semi-major axis ($a$) of the bounding mass spheroid, the power-law index of the particular mass model (i.e. Ferrers or Hernquist), and the inclination angle ($i$) of the symmetry axis (i.e. either $i = 90°$ or $i =$ true inclination of cluster). Then for a given ellipticity of the dark matter ($\epsilon_{DM}$; really the total mass but since the dark matter dominates the potential we refer to it as the dark matter) we generate model X-ray surface brightness maps for any values of the free parameters $R_c$ and $\Gamma$ associated with the particular solution of the hydrostatic equation; i.e. isothermal or arbitrary temperature profile. The procedure is the same for the SP models except that (1) the boundary of the mass is not specified, and (2) the power-law index of the potential (logarithmic or $n = 0.1 - 0.5$) and the ellipticity of the potential ($\epsilon_\Phi$) are specified. We determine the free parameters by performing a $\chi^2$ fit that compares the radial profiles of the model and image. Since the images from the simulation have essentially no noise (see §2.), the best-fit parameters are taken to be the only solution. We determine a particular model to be consistent with the image if $\epsilon_M$ computed from the model in either the $5' - 20'$ or $5' - 40'$ apertures is consistent with that computed from the image in

§2.1.. Those models that are consistent with neither are rejected. We take the union instead of the intersection because the ellipticity in one of the apertures may be affected by clumping at a particular radius or by incomplete relaxation of the gas. Thus we aim to include all of the models consistent with the aggregate shape of the cluster on scales of $\sim 1.5$ Mpc.

Recall that we want to compare the aggregate shape of the true dark matter from the simulation with the dark matter from the models on a scale of $\sim 1.5$ Mpc; i.e. a comparison of their quadrupole moments, or equivalently, their principal moments of inertia. The SMD models have dark matter that is of constant ellipticity and thus the aggregate shape is the same as that computed on smaller scales. For the SP models, however, the dark matter changes shape with radius and thus we employ the iterative moment technique (see §2.1.) to obtain the desired aggregate ellipticity.

We display in Figure 7 the results for the ellipticity of the dark matter for both the SMD and SP models; the true dark matter ellipticity computed within 1.5 Mpc ($\epsilon_{DM} = 0.54$, see §3.) is represented by a horizontal dashed line in the Figure 7. For the models where the true inclination of the symmetry axis is used the ellipticities of the dark matter models agree well with the true value from the simulation. The true-temperature models generally agree within $\Delta\epsilon \sim 0.05$ of the true value while the isothermal models underestimate the ellipticity by $\Delta\epsilon \sim 0.10$; these deviations are within the typical estimated errors obtained by BC92 for *Einstein* IPC clusters using the SMD models. Note that due to the projection properties of oblate and prolate spheroids (e.g., Fabricant et al. 1984) the oblate models for the $y$ and $z$ models corrected for the true cluster inclination represent essentially the upper halves of the ellipticity ranges in Figure 7 while the prolate models correspond to the lower halves; i.e. the oblate models give better agreement for these cases. As expected, when the inclination of the symmetry axis is not taken into account the deviations from the true ellipticity increase considerably. For the $y$-projection the true-temperature models underestimate the true dark matter ellipticity by $\sim 0.10$ and the isothermal models by $\sim 0.20$; note that the former deviation is within the typical uncertainty of BC92. For the $z$-projection the deviations are $\sim 0.25$ and $\sim 0.35$ respectively. The difference in ellipticity between the true-temperature and isothermal models, however, is typically $\sim 0.10$ which is comparable to the uncertainty of BC92.

Similar to the $\beta$-models (see §2.1.), the best-fit models are generally excellent visual fits to the radial profile of the surface brightness for most of the models considered. The fits do not distinguish between oblate and prolate models, consistent with the real triaxiality of the dark matter (see §3.). As reflected by their $\chi^2$ values, the SMD models having $\rho_{DM} \sim r^{-2}$ (and SP logarithmic models) generally fit the simulation data better than the $\rho_{DM} \sim r^{-3}$ (and SP $n = 0.5$) models, although visually the differences are not flagrant. The Hernquist models fit the data well but with large core parameters indicating that the $r^{-4}$ regime is suppressed. This behavior is consistent with that of the true dark matter (§3.). The semi-major axis of the SMD models (set to 1.75 Mpc) is not well constrained and the dark matter shapes are not very sensitive to it. However, the quality of the fits diminishes for smaller $a$ and we estimate a lower limit $a > 0.5$ Mpc from visual examination of the fits.

The core parameters $R_c$ of the models behave differently for the isothermal and true-temperature models. For an isothermal gas the equation of hydrostatic equilibrium requires the core radius of the gas density and the total gravitating mass to be nearly the same; our models reproduce this expected similarity. For the true-temperature models we obtain model core radii in excellent agreement with that of the true dark matter core radii (see §3.).

### 4.3. Results for $z > 0.13$ Clusters

Recall from §3. that the gas isodensity surfaces for the $z = 0.83, 0.67$ and $z = 0.38$ clusters do not trace the isopotential surfaces. The X-ray emitting gas in the $z = 0.83$ nearly traces the dark matter itself, not the potential while for the $z = 0.67$ and $z = 0.38$ clusters the X-rays trace neither the dark matter nor the potential. In fact, the distortion of the isophotes of the $z = 0.38$ cluster suggests the gas is "sloshing" as a result of the infall of the clump seen in the upper-right of the earlier redshift plots. These properties suggest that the gas is out of equilibrium and the hydrostatic analysis of the mass distribution of these clusters is not justified. Indeed large errors in the derived dark matter shapes result; e.g., the ellipticity of the dark matter derived for the $z = 0.38$ cluster is less than the true ellipticity by greater than 0.25 for all models considered. The $z = 0.25$ cluster, in contrast, mirrors the $z = 0.13$ case by giving excellent agreement between the X-ray-derived shapes and the true dark matter shapes; i.e. we obtain $\epsilon_{DM} = 0.46 - 0.55$ for the true temperature models and $\epsilon_{DM} = 0.40 - 0.49$ for the isothermal models, comparable to the true dark matter ellipticity of $\epsilon = 0.55$.

### 5. Discussion

Do the peculiar features (i.e. biased CDM, no star formation) of the KW simulation preclude generalizing the results of the previous section to real clusters? The primary virtue of the KW simulation is that it produces a "non-trivial" cluster: the KW cluster (1) is quite flattened having an ellipticity of about 0.55 within 1.5 Mpc, (2) has a steep temperature gradient that does not appear to be typical of real clusters (Mushotzky 1994), and (3) has substructure at all redshifts. Surely if the KW simulation produced a round, isothermal, and smooth cluster the X-ray methods could not have failed. Therefore, the KW cluster may not be a perfect representation of a real cluster but it provides a formidable test for the X-ray methods of shape determination.

We determined for the KW cluster that the X-ray method for constraining the aggregate shape of the dark matter on a scale of $r \sim 1.5$ Mpc is valid for $z \lesssim 0.25$. If, however, the evolution of a real cluster substantially differs from the KW cluster (e.g., because of a different cosmology or the presence of star formation) then these "safe" redshifts for X-ray analysis may not apply to a real cluster; for a discussion of the effect of cosmology on the epoch of cluster formation see, e.g., White (1994).

The X-ray images of the KW cluster exhibit general properties as a function of redshift that correlate with the reliability of the X-ray methods. Clearly the strong subclustering in the $z = 0.83, 0.67$ and $z = 0.38$ clusters and the distorted X-ray isophotes in the $z = 0.38$ cluster are not seen at the lower redshifts. Moreover, the isophotes of the lower redshift clusters ($z \lesssim 0.25$) are overall more regularly shaped and rounder than those at higher redshifts. If the gas was in hydrostatic equilibrium at the earlier times ($z \gtrsim 0.38$) then the large ellipticities ($\epsilon_x \gtrsim 0.4$) of their isophotes would imply dark matter ellipticities larger than 0.7 (cf. end of §5.1 of BC92); this is unphysical because dynamical considerations forbid such flat, non-rotating, ellipsoidal structures (Merritt & Stiavelli 1990; Merritt & Hernquist 1991). Thus a qualitative statement of necessary conditions for the reliability of the X-ray methods is that (1) there is no obvious subclustering on the same scale used to compute the aggregate shape and (2) the isophotes are regularly shaped and not too elongated ($\epsilon_x \lesssim 0.3$). Of course these conditions are not sufficient since they could both be the results of projection effects.

The clusters studied by BC92 satisfy these necessary conditions with the possible exception of Coma. Fitchett & Webster (1987) have suggested that Coma is bimodal on scales of several hundred kpc, comparable to the scale used by BC92 to compute the aggregate shape. Davis & Mushotzky (1993) have provided further evidence for such bimodality from analysis of *Einstein* X-ray data. For A2256, in contrast, the substructure appears to reside in the core on a scale of a few hundred kpc (e.g., Briel et al. 1991) which is substantially smaller than the aggregate scales ($\sim 1$ Mpc) used by Fabricant et al. (1984) and Buote (1992) to measure the intrinsic dark matter shape. Moreover, the core substructure in A2256 appears to be very similar to that present in the $z = 0.13$ cluster of the KW simulation. *We thus conclude that core substructure representing scales of a few hundred kpc does not invalidate X-ray measurements of intrinsic aggregate (i.e. $r \sim 1 - 2$ Mpc) cluster shapes.*

Although we have examined simulated X-ray images having essentially unlimited photon statistics, the additional uncertainties due to noise for the Abell clusters studied by BC92 with *Einstein* are not prohibitive (see Table 7). However, a large sample of such clusters is required for intrinsic shapes of clusters to be employed as a cosmological constraint (see §1.). In order to obtain shape constraints with the ROSAT PSPC of similar quality to BC92 we need to restrict ourselves to clusters that are sufficiently bright (for $S/N$) and nearby (for sufficient angular resolution). From examination of the ROSAT master log of pointed observations (in the HEASARC-Legacy data base) for Abell clusters having (1) a measured flux $> 10^{-11}$ erg cm$^{-2}$ s$^{-1}$ as published by Ebeling (1993), (2) exposure times $\geq 5000$s, and (3) $z < 0.11$, we find 36 eligible clusters. Higher redshift clusters will be available for analysis with AXAF because of its superior spatial resolution (FWHM $1.9''$ at 1 keV). As a result, a total of 124 Abell clusters from Ebeling (1993) having flux $> 10^{-11}$ erg cm$^{-2}$ s$^{-1}$ will in principle be eligible for analysis. It is difficult to interpret these numbers of eligible clusters because we do not know how many have substantial substructure that invalidates the X-ray shape analysis. However, our analysis of the KW simulation demonstrates that the presence of core substructure does not invalidate the shape analysis thus indicating that a sizeable fraction of the 36 PSPC and 124 AXAF Abell clusters should enable reliable X-ray constraints of their intrinsic shapes.

## 6. Conclusions

We investigate the reliability of X-ray methods for determining the intrinsic shapes of galaxy clusters by analyzing the cluster simulation of Katz & White (1993); the effects of subclustering and temperature gradients on the shape determinations are examined. Specifically, we test the X-ray technique used by BC92 (BC94; Buote 1992), who built on the original study of Binney & Strimple (1978; Strimple & Binney 1979), to constrain the shapes of the dark matter in five Abell clusters using *Einstein* images. In order to reduce effects of small-scale substructure (few hundred kpc) we measure the aggregate shapes of clusters on scales of $1 - 2$ Mpc from the cluster center.

For low redshifts ($z \lesssim 0.25$) we find that the X-ray method accurately measures the true ellipticity of the cluster dark matter when the true inclination of the cluster is taken into account. The X-ray models employing the true temperature profile deviate from the true cluster ellipticity ($\epsilon \sim 0.55$) by $\epsilon \sim 0.05$ while the isothermal models have slightly larger deviations $\epsilon \sim 0.10$; both of these deviations underestimate the true ellipticity but are less than the typical uncertainties obtained by BC92 for real clusters.

The reason for this underestimate is the following. The hydrostatic equation requires that the gravitating mass has a core radius similar to that of the gas itself when the gas is isothermal, but it has a smaller core radius when the gas has a negative temperature gradient. The core radius of the gas for the KW cluster is about five times larger than the core radius of the dark matter (see §3.); i.e. the isothermal solution for the KW cluster is less centrally condensed than the true-temperature solution. At a given radius the spherically-symmetric monopole term in the gravitational potential is more important for the centrally-condensed cluster. Thus, in order to generate the same ellipticity (i.e. quadrupole) of the potential at a given distance, the cluster having a negative temperature gradient for the X-ray emitting gas will have to be more elongated than the cluster having an isothermal gas.

When the inclination of the cluster is not taken into account we obtain results for the true temperature models in accordance with Binney & Strimple (1979) and Fabricant et al. (1984); of course, the effects of inclination on cluster shapes may be uncovered by analyzing a well-defined statistical sample of clusters (e.g., Plionis, Barrow, & Frenk 1991). Our results affirm the assertion that conclusions regarding the shape of the dark matter are not overly sensitive to the temperature gradient of the gas (§4.1.); i.e. the ellipticities of the true-temperature models differ from the isothermal models by less than the typical statistical uncertainties of BC92. We expect that the assumption is even more valid for real clusters since they likely do not have such a steep temperature gradient like that present in the simulation (§3.).

At higher redshifts ($0.38 \lesssim z \lesssim 0.83$) the X-ray method yields unreliable results. The gas at these early times does not trace the shape of the cluster gravitational potential as it must if it were in hydrostatic equilibrium. At $z \sim 0.83$ the gas traces the dark matter itself and for $z \sim 0.38 - 0.67$ it follows neither the dark matter nor the potential. Since the peculiarities of the simulation (§2.) may obfuscate interpretation of the results at these redshifts in terms of real clusters, we offer qualitative necessary conditions for the reliability of X-ray methods characterized by both the amount of substructure in the X-ray surface brightness and the shapes of the X-ray isophotes.

We conclude that measurements of the aggregate shapes of clusters on scales of $1 - 2$ Mpc from the cluster center are practically unaffected by core substructure representing scales of a few hundred kpc. Therefore our results suggest that the X-ray studies of such aggregate shapes of clusters by Fabricant et al. (1984) and BC92 (Buote 1992) are valid provided that they do not suffer from serious projection effects.

Since our analysis of the KW simulation demonstrates that the presence of core substructure does not necessarily invalidate X-ray shape analysis, we conclude that a sizeable fraction of 36 bright, low-redshift ($z \lesssim 0.1$) Abell clusters from ROSAT PSPC pointed observations should yield reliable X-ray constraints of their intrinsic shapes. With the inclusion of higher redshift ($z \lesssim 0.3$) clusters, AXAF can observe in principle 124 candidate Abell clusters.

We are grateful to Neal Katz for graciously allowing us to use the KW simulation. DAB acknowledges Claude Canizares and Eric Gaidos for useful discussions. We thank Claude Canizares for a critical reading of the manuscript. We extend our appreciation to Timothy Beers for his prompt refereeing of this paper and for his useful suggestions. DAB acknowledges grants NAS8-38249 and NASGW-2681 (through subcontract SVSV2-62002 from the Smithsonian Astrophysical Observatory). JCT was supported by an NRC associateship.

Table 1: X-ray Ellipticities and Position Angles

| $a$ | | $z = 0.13$ | | | | | | $z = 0.25$ | |
| (arcmin) | (kpc) | $x$ | | $y$ | | $z$ | | $x$ | |
| | | $\epsilon_M$ | $\theta_M$ | $\epsilon_M$ | $\theta_M$ | $\epsilon_M$ | $\theta_M$ | $\epsilon_M$ | $\theta_M$ |
|---|---|---|---|---|---|---|---|---|---|
| 10 | 291 | 0.26 | 20 | 0.19 | 10 | 0.04 | 27 | 0.17 | 49 |
| 15 | 436 | 0.26 | 15 | 0.16 | 11 | 0.04 | 20 | 0.17 | 46 |
| 20 | 582 | 0.27 | 18 | 0.16 | 7 | 0.07 | 19 | 0.17 | 46 |
| 25 | 728 | 0.27 | 20 | 0.15 | 5 | 0.08 | 20 | 0.17 | 46 |
| 30 | 873 | 0.27 | 21 | 0.14 | 2 | 0.08 | 20 | 0.18 | 44 |
| 35 | 1018 | 0.25 | 22 | 0.13 | 0 | 0.08 | 20 | 0.19 | 44 |
| 40 | 1164 | 0.25 | 23 | 0.13 | 0 | 0.09 | 20 | 0.20 | 44 |

Note. — $a$ is the semi-major axis of the elliptical aperture used to compute the iterative moments (see §2.1.); the inner $5'$ is not included so the aperture is actually the annulus defined from $5' - a$. $\theta_M$ is in degrees measured with respect to the horizontal axis in Figure 1.

Table 2: True 3-D Dark Matter and Gas Shapes for $z = 0.13$ Cluster

| $a$ | | Dark Matter | | | Gas | | |
| (arcmin) | (kpc) | $\epsilon_{21}$ | $\epsilon_{31}$ | $\theta_{DM}$ | $\epsilon_{21}$ | $\epsilon_{31}$ | $\theta_{gas}$ |
|---|---|---|---|---|---|---|---|
| 10 | 291 | 0.15 | 0.30 | 0 | 0.09 | 0.27 | 30 |
| 15 | 436 | 0.11 | 0.35 | 5 | 0.06 | 0.25 | 16 |
| 20 | 582 | 0.04 | 0.39 | 11 | 0.12 | 0.29 | 12 |
| 25 | 728 | 0.10 | 0.41 | 9 | 0.12 | 0.30 | 8 |
| 30 | 873 | 0.08 | 0.45 | 9 | 0.12 | 0.33 | 7 |
| 35 | 1018 | 0.12 | 0.45 | 9 | 0.13 | 0.34 | 10 |
| 40 | 1164 | 0.18 | 0.49 | 8 | 0.12 | 0.31 | 9 |
| 45 | 1310 | 0.21 | 0.52 | 8 | | | |
| 50 | 1455 | 0.27 | 0.54 | 9 | | | |

Note. — $a$ is the semi-major axis of the ellipsoidal aperture used to compute the iterative moments (see §3.); the inner $5'$ is not included in the gas so the aperture in that case is actually the annulus defined from $5' - a$. $\epsilon_{21}$ is the ellipticity in the plane of the two largest principal moments and $\epsilon_{31}$ is the ellipticity in the smallest-largest principal moment frame. The position angles are relative positions of the smallest principal axes ($I_3$): $\theta_{DM}$ is the relative direction of the dark matter at semi-major axis $a$ with respect to $a = 10'$; $\theta_{gas}$ is the relative direction of the gas at $a$ and the dark matter at $a$.

Fig. 1.—
Contour plots of the X-ray surface brightness of the KW cluster (see §2.1.) at $z = 0.13$ for the three orthogonal projections $x$ ($i = 80°$), $y$ ($i = 60°$), $z$ ($i = 40°$); the contours are separated by a factor of 2 in intensity and the coordinate axes are labeled in arcminutes. The cluster is placed at a distance of 100 Mpc so that $1'$ represents 29.1 kpc.

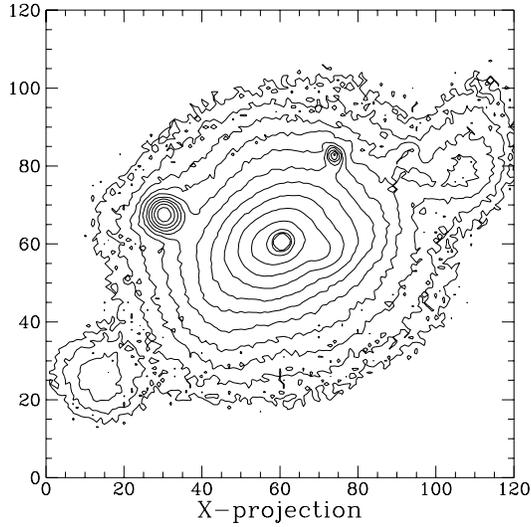

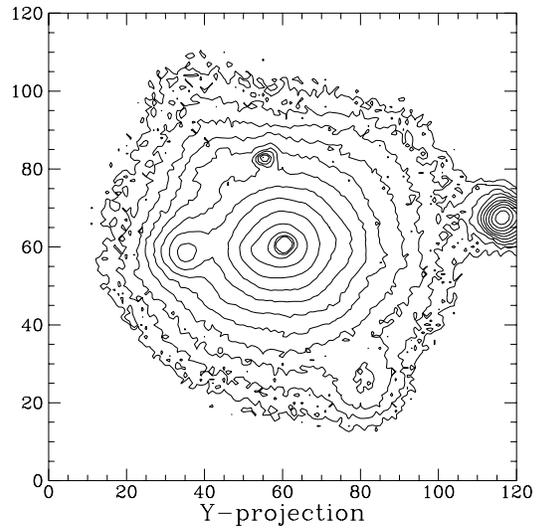

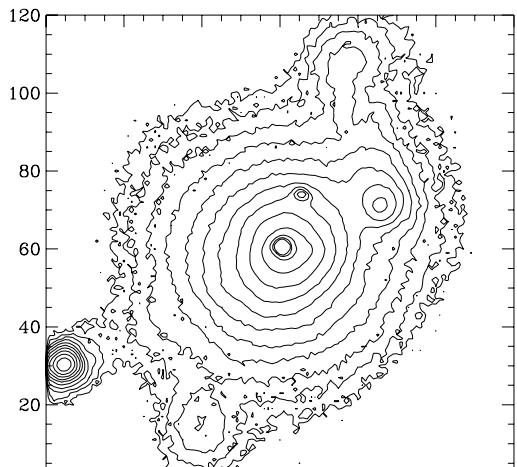

Fig. 2.—
Contour plots of the same images in Figure 1 where now the images have been corrected for
contamination from individual globs (see§2.1.).

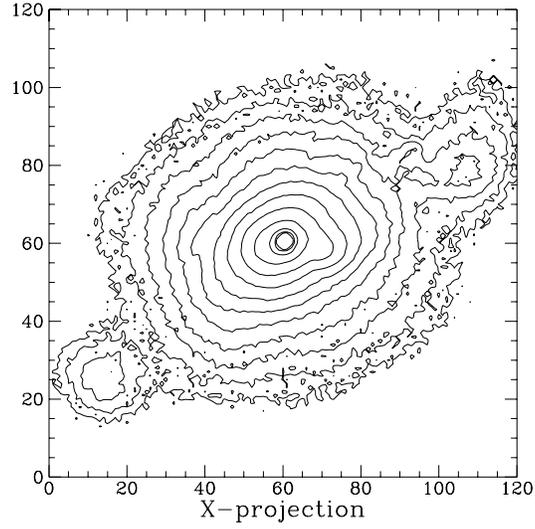

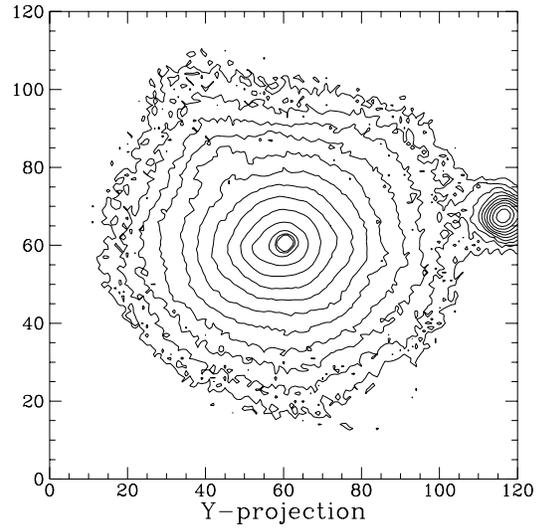

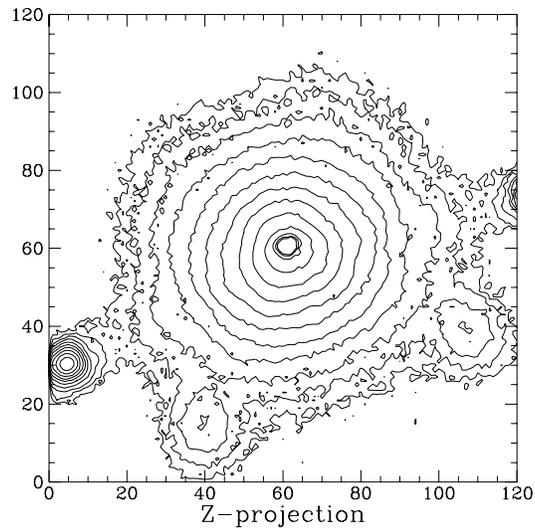

Fig. 3.—
The azimuthally-averaged radial profiles of the reduced $z = 0.13$ images from Figure 2 and their best-fit $\beta$ models.

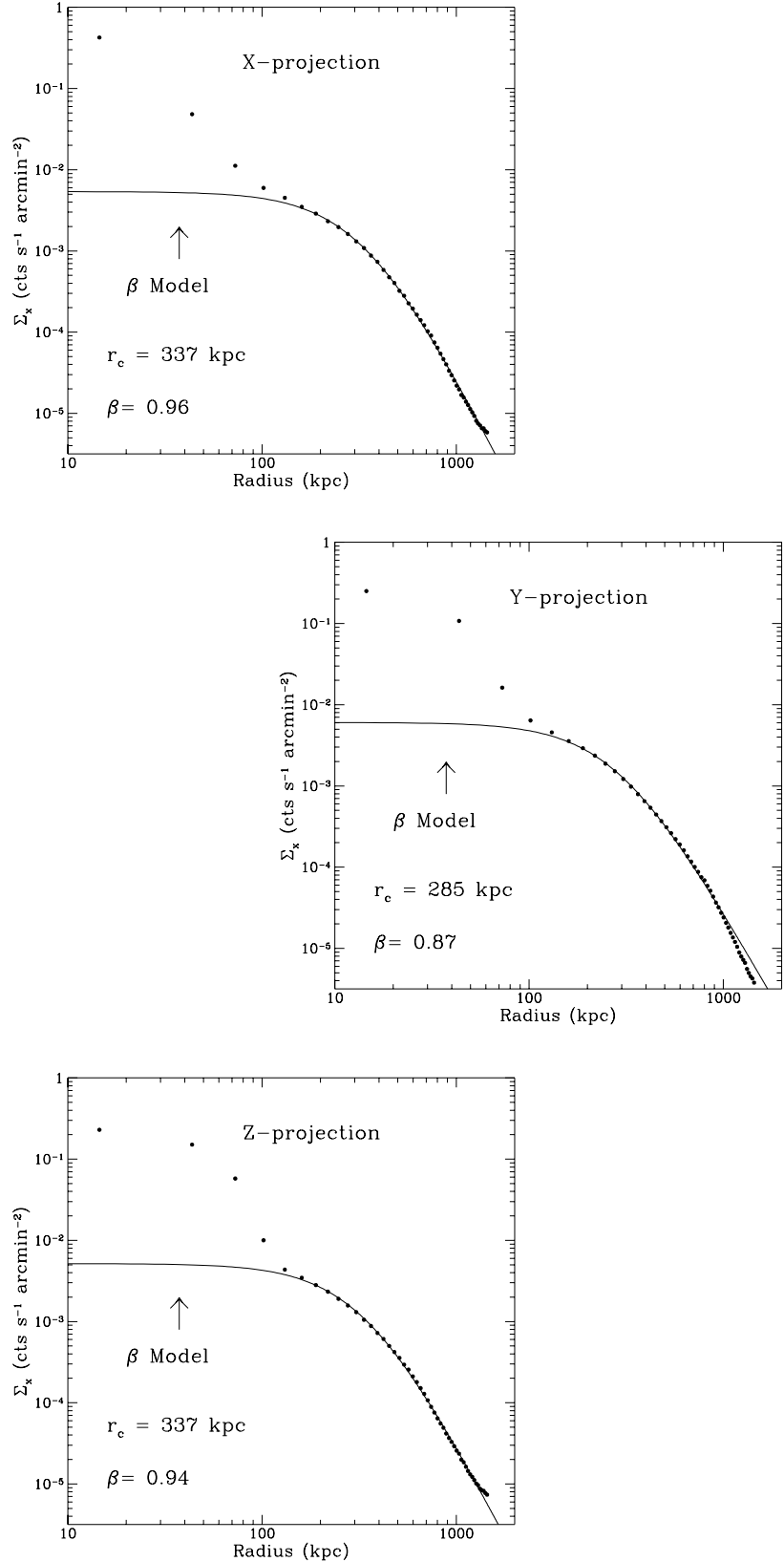

Fig. 4.—
Contour plots of the X-ray surface brightness for the $x$-projection of the KW cluster at higher redshifts (see§2.2.); the images are prepared as in Figure 1.

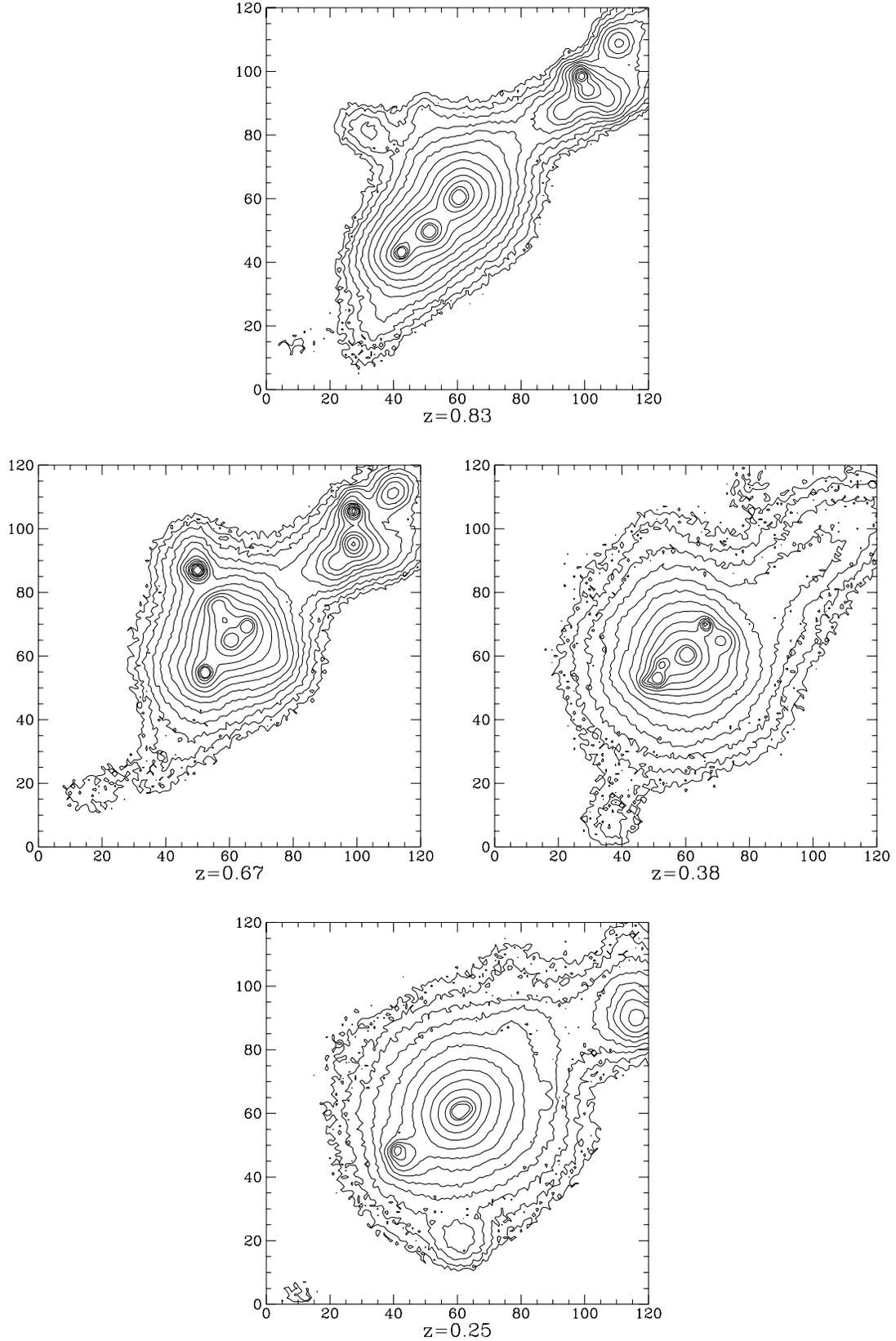

Fig. 5.—
(a) Contour plot of the $z = 0.25$ image in Figure 4 where now the image has been corrected for contamination from individual globs (see §2.1. and §2.2.).

(b) The azimuthally-averaged radial profile of the reduced $z = 0.25$ image and the best-fit $\beta$ model.

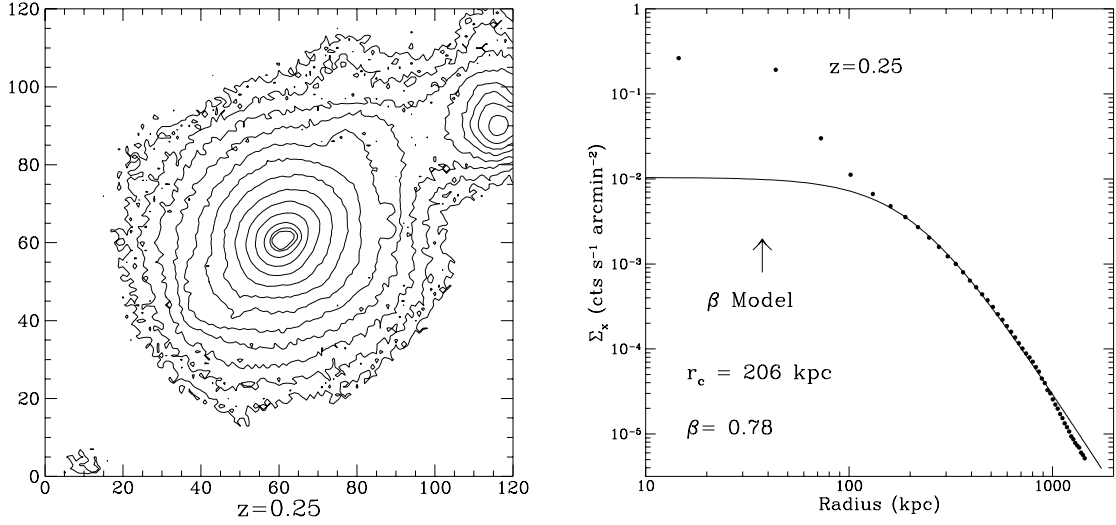

Fig. 6.—
Contour plot of the $z = 0.13$ dark matter projected in the plane of the longest and shortest principal moments of inertia; the plot has been smoothed with a Gaussian filter ($\sigma = 3'$ FWHM) for visual clarity and the contours a separated by a factor of two in mass. See Table 2 for the ellipticity and degree of triaxiality of the dark matter.

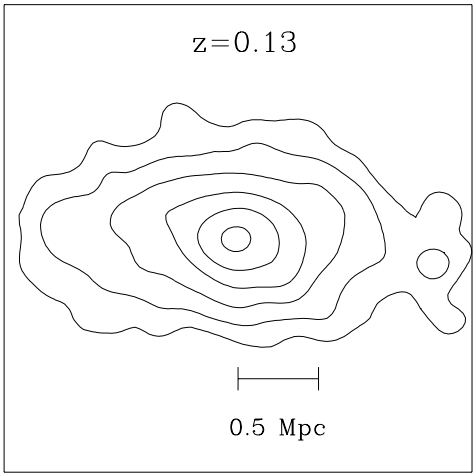

Fig. 7.—
Results for the intrinsic three-dimensional ellipticity of the dark matter of the $z = 0.13$ cluster as
estimated from the spheroidal X-ray models in §4.1.. The ellipticities represent the aggregate
shape of the dark matter computed for $r \lesssim 1.5$ Mpc (see §3. and §4.2.). The long dashed line
represents the aggregate dark matter ellipticity of 0.54 computed directly from the simulation in
§3.. The solid error bars represent the range of X-ray models where the true temperature profile
of the simulation has been used while the dotted error bars indicate the results for the isothermal
models. The models enclosed in boxes have been corrected for the true inclination of the cluster.
The dumbbell represents the typical uncertainty obtained by BC92 for the dark matter
ellipticities of five Abell clusters with the *Einstein* IPC.

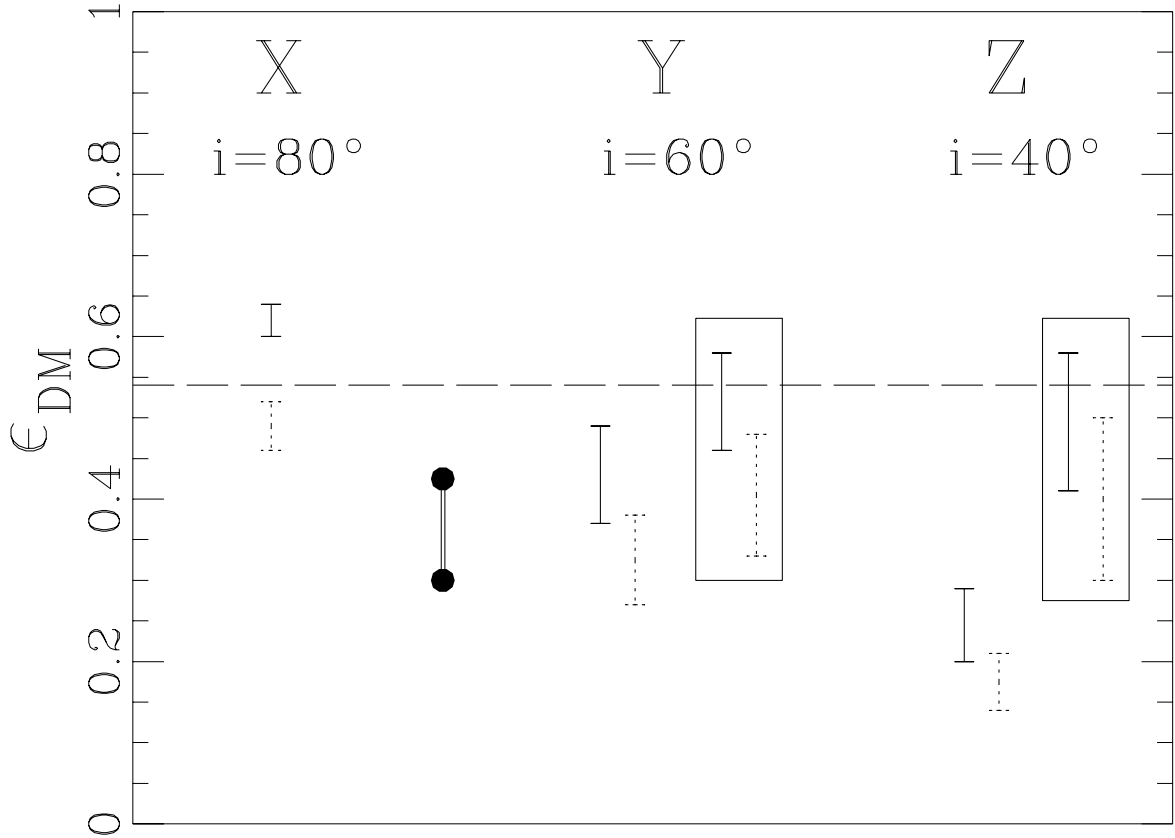